\documentclass{article}
\usepackage[utf8]{inputenc}
\usepackage{amsmath}
\usepackage{amssymb}
\usepackage{amsfonts}
\usepackage{enumitem}
\usepackage{hyperref}
\usepackage{xcolor}
\usepackage{bbold}
\usepackage{bm}
\usepackage{graphicx}
\usepackage{float}
\usepackage[margin=1.5in]{geometry}
\usepackage{titling}
\usepackage{graphics}
\usepackage{caption}
\usepackage{subcaption}

\title{Evolutionary Strategies with Analogy Partitions in $p$-guessing Games}
\author{Aymeric Vi\'{e}\\Paris School of Economics, Paris, France\\New England Complex Systems Institute, Cambridge, MA, USA}

\begin{document}


\maketitle

\section*{Abstract}
In Keynesian Beauty Contests notably modeled by $p$-guessing games, players try to guess the average of guesses multiplied by $p$. Theoretical and experimental research in the spirit of level $k$ models has characterized the behavior of agents with different levels of reasoning when $p$ is persistently above or below $1$. Convergence of plays to Nash equilibrium has often been justified by agents' learning. However, interrogations remain on the origin of reasoning types and equilibrium behavior when learning takes place in unstable environments. When successive values of $p$ can take values above and below $1$, bounded rational agents may learn about their environment through simplified representations of the game, reasoning with analogies and constructing expectations about the behavior of other players. We introduce an evolutionary process of learning to investigate the dynamics of learning and the resulting optimal strategies in unstable $p$-guessing games environments with analogy partitions. As a validation of the approach, we first show that our genetic algorithm behaves consistently with previous results in persistent environments, converging to the Nash equilibrium. We characterize strategic behavior in mixed regimes with unstable values of $p$. Varying the number of iterations given to the genetic algorithm to learn about the game replicates the behavior of agents with different levels of reasoning of the level $k$ approach. This evolutionary process hence proposes a learning foundation for endogenizing existence and transitions between levels of reasoning in cognitive hierarchy models. \\

\textbf{Keywords}: Analogy reasoning, Genetic algorithm, Level-$k$ model, $p$-guessing games.\\

\textbf{JEL codes}: C63, C73, D83 

\clearpage
\section{Introduction}

Consider a game in which players announce a number in a given interval. The player with the announcement closest to the average of announcements multiplied by $p$ wins. The positive parameter $p$ is set to be common knowledge among players, and is predetermined. $p$-guessing games initially introduced by \textcolor{blue}{Moulin (1986)} have often been linked to investigations on the behavior of players, and their beliefs dynamics, leading to the well known level $k$ model \textcolor{blue}{(Nagel, 1995)}. Players are assumed to differ in their sophistication by endowment of different degree of reasoning depth. Level $0$ players, least sophisticated, are assumed to select a strategy at random in the interval of possible announcements. A more sophisticated level $1$ player best responds to the belief that all players are level $0$, forming first-order beliefs on their behavior. An even more sophisticated level $2$ agent forms second-order beliefs about the behavior of other players assumed to be of level $1$. This characterization of the degree of reasoning of players up to the $n$-level goes back to \textcolor{blue}{Keynes (1936)} who famously wrote: ``\textit{It is not a case of choosing those [faces] that, to the best of one's judgment, are really the prettiest, nor even those that average opinion genuinely thinks the prettiest. We have reached the third degree where we devote our intelligence to anticipating what average opinion expects the average opinion to be. And there are some, I believe, who practice the fourth, fifth and higher degrees.}''. \\

Game theoretical and experimental approaches alike have devoted a substantial attention to the study of these games, in relation to phenomena of beauty contests in the formation of expectations and determination of economic behavior. For $0 \leq p < 1$, there exists a unique Nash equilibrium in which players play $0$, by iterated elimination of dominated strategies. When $p > 1$, no dominated strategies exist and the game admits infinitely many equilibrium points in which all players choose the same number, as the game becomes a coordination issue (\textcolor{blue}{Jack Ochs, 1995}). Equilibrium emerges as a result of infinite level of reasoning, or the presence of focal points for coordination, such as the ``Schelling salience'' (\textcolor{blue}{Schelling, 1980}). Various experimental studies (\textcolor{blue}{Nagel, 1995; Duffy and Nagel, 1997}) outlined that while in opening rounds, outcomes were quite divergent from Nash equilibrium, subsequent rounds showed much closer outcomes to the predictions of the theory. Learning models in which the agents acquire understanding of the game and the strategy environment have been popular to explain this convergence to equilibrium (\textcolor{blue}{Stahl, 1996; Weber, 2003}) without relying on improperly high levels of reasoning. Subjects were often found to identify depths of reasoning generally between order $0$ and $2$ (\textcolor{blue}{Duffy and Nagel, 1997}). The behavior of players in first stages of the game was examined to be consistent with the learning direction theory of \textcolor{blue}{Selten and Stoecker (1986)} and \textcolor{blue}{Nagel (1995)} by (\textcolor{blue}{Duffy and Nagel, 1997}), an ex-post reasoning process in which announced winning numbers provoke an increase, or decrease of the depth of reasoning. Individuals adjust in reasoning depth in this view to the winning number, as a reference point. Non-winning players were experimentally  similarly shown to be imitating the level of rationality of winners in \textcolor{blue}{Sbriglia (2008)}, imitation and best subsequent response to imitative strategies increasing substantially the learning process.\\

Can we achieve this convergence result when agents face changing environments to learn their strategies? While learning in persistent environments typically makes subjects' actions converge to the Nash equilibrium (\textcolor{blue}{Duffy and Nagel, 1997}), no attention has been given to strategic behavior and learning when economic agents face successively games where $p < 1$ and $p > 1$. In such complicated environments, agents may form simplified representatons of the world, form expectations about the game parameter and the behavior of other players. A thorough framework building on this intuition has been proposed by (\textcolor{blue}{Jehiel, 1995}) exploring analogy-based expectation equilibrium, where learning ocurrs through the formation in agents' minds of analogy partitions. Consider an extension of the $p$-guessing game where with probability $q$, $p$ is distributed uniformly on $(0,1)$, and with probability $1-q$, $p$ is distributed uniformly on $(1,2)$. Assume that agents learn through repeated play an optimal strategy, bundling experienced games together. In complex games, agents may use simplified representations of the world to understand its behavior, and develop strategies that best respond to this expectation backed by experiences on the average behavior (\textcolor{blue}{Jehiel, 1995}). When bounded rational agents evolve in a repetition of $p$-guessing where the parameter $p$ is no longer predetermined across all repetitions and common knowledge, but a random variable distributed with a distribution that players learn by experience, what strategies are played in equilibrium as a result of learning?\\

In such a situation, strategy spaces are large, making traditional methods of investigation hard to employ, and equilibrium strategies difficult to identify. On the other hand, recent tools from computer science have shown ability to thrive in environments of high strategical complexity such as go, chess, shogi and wargames (\textcolor{blue}{Silver et al. (2016, 2017, 2018), Vinyals et al., 2019}). To be able to characterize equilibrium strategies in $p$-guessing games with learning with analogy partitions, we propose to rely on a genetic algorithm, inspired by the recent progress of combinations of algorithmic efficiency and games. This popular evolutionary process well presented by (\textcolor{blue}{Mitchell, 1998}) is consistent with viewing bounded rational players finding optimal strategies by learning, or fictions play: strategies are played against different expectations of the behavior of other agents themselves learning with the same process. Strategies' outcomes are compared, the best are selected and recombined to create the next generation of strategies. Mutations (peryurbations) allow to explore deviations from these strategies, and cover a sufficient diversity in the search process. The use of genetic algorithm, and more boradly of evolutionary methods in economics is not new (\textcolor{blue}{Friedman, 1998}). Its relevance for learning in evolutionary game has been highlighted notably by \textcolor{blue}{Reichman, 2001}, and for its interest to model bounded rationality and learning (\textcolor{blue}{Arthur, 1994, Edmonds and Moss, 1997}) notably applied in the context of game theory (\textcolor{blue}{Marks, 2002}). The genetic algorithm does not consider all possible strategies before making its choice, but evaluates the performance of its initial candidate strategies, combine the best to form new ones, and locally perturbs these strategies to explore alternatives. It is consistent with a view of learning with bounded rational agents, that offers strong synergies with experimental methods (\textcolor{blue}{Duffy, 2006}).\\

We introduce a genetic algorithm to investigate the dynamics of learning and the resulting optimal strategies in such environments with analogy partitions. As a validation of the approach, we first show that our genetic algorithm behaves consistently with theoretical and experimental results in persistent environments with fixed $p$, converging to the Nash equilibrium. We explore strategic behavior in $p$-guessing games when agents' learning proceeds in unstable environments, when the state of nature, i.e. the value of $p$ higher or inferior to $1$ follows a given probability in the interior of the interval $(0,1)$. Varying the number of iterations given to the genetic algorithm to learn about the game replicates the behavior of agents with different levels of depth of reasoning in the level $k$ approach. This evolutionary process hence proposes a learning foundation for endogenizing existence and transitions between levels of reasoning in cognitive hierarchy models. Section \ref{algorithm} introduces the evolutionary algorithm specifications and components, as well as the model of the game. Our genetic algorithm calibration procedure is described in section \ref{calibration}. Results on the convergence of learning with analogy partitions in persistent as well as unstable environments are presented in section \ref{results}. Section \ref{discussion} opens on the possibility to endogenize the existence and transitions between levels of reasoning depths in a learning interpretation. Section \ref{conclusion} concludes.

\section{Learning algorithm}
\label{algorithm}

\subsection{$p$-guessing games in unstable environments}

We consider the following game. Players attempt to guess the average of other players' guesses multiplied by a positive scalar $p$. $p$ is determined according to the probability $q$. With probability $q$, $p$ is uniformly drawn in $(0,1)$. We consider an environment in which actions, or announcements, are in the interval $[0,10]$. With probability $1-q$, $p$ is drawn from a uniform distribution $(1,2)$. Our learning process places a single-agent in repeated games, where the value of $p$ is determined by repeated draws of $q$. Agent's strategies are compared to beliefs about the average behavior, analog to the behavior of other players endowed with the same learning process. These beliefs, and the evolution of strategies, are build by bundling previous repetitions of the game together in an analogy class. 

\subsection{Initial strategies}

We start the learning process by considering a vector of strategies denoted \textbf{$\textbf{S}_{t=0}$}. It is composed at the initialization of the process of integers covering the strategy space of announcements $(0,10)$, in order to cover enough diversity of the strategy space at the start. Our agent does not consider all possible strategies in this interval, but is given the ability to evaluate and choose diverse strategies within this range. For simplicity, let us assume that $\textbf{S}_{t=0} = (0, 1, \dots, 9, 10)$. 

To this vector of strategies, at each iteration $t$, we associate a vector of payoffs $\textbf{P}_t$ of same dimension, corresponding to the payoff our agent would obtain upon playing each possible strategy, that is, each element of \textbf{$\textbf{S}_{t}$}.

\subsection{Evaluation of strategy dominance}

Once the game environment has been drawn, each candidate strategy is compared with respect to the payoff it yields. At each iteration $t$, each strategy -i.e. each element of $\textbf{S}_{t}$- is played against each other elements of $\textbf{S}_{t}$. In our evolutionary approach, our player plays the games within its mind, testing each strategy against each other strategy, corresponding in this process to beliefs about the average behavior of other players. As an illustration, consider $\textbf{S}_{t=0} = (0, 1, \dots, 9, 10)$. In evaluating the payoff of announcing $0$, the agent will compute his utility from playing $0$ when the average behavior of other players is to announce $1$, then $2$, and so on until $10$. The resulting sum of payoffs gives the average payoff from playing action $0$. This evaluation process is iterated for all elements of $\textbf{S}_{t=0}$ in our illustration. To avoid indifference between alternatives, and ensure adequate differentiation of strategies based on their performance, we assume in this process that payoffs $u(x)$ are characterized by a continuous loss function. Defining $x$ as the action being played, $p$ the game parameter and $\overline{x}$ the average behavior of other players, we obtain:

$$ u(x) = - (x - p\overline{x})^2$$

The utility attached to each strategy is a decreasing function of the distance between the strategy played, and the average behavior of other players multiplied by the parameter $p$ in our $p$-guessing game. This measure $u(x))$ evaluated for all elements $x$ allows to identify dominating as well as dominated strategies in the strategy vector $\textbf{S}_{t}$ at iteration $t$. Using utility values as criterion, best strategies are selected are recombined to form new strategies. Local perturbations denoted \textit{mutations} allow for exploration with strategy restrictions of the space of possible strategies, ensuring overall improvement throughout iterations of the strategy pool.

\subsection{Genetic Algorithm and creation of new strategies}

There exist several methods to select parents strategies in a genetic algorithm, responding to a necessary trade-off between diversity of the solutions, and their fitness. The objective is to improve the utility of the set of strategies, while maintaining enough ``genetic'' diversity to explore the space of possible actions. We adopt in this model a \textit{tournament method}. An arbitrary number of $3$ strategies is selected in the strategy space, their net payoffs compared. The resulting two best are selected to be the parents of a new strategy. We proceed by the \textit{uniform crossover} method: the resulting ``child'' strategy is the average of the two parents selected.\\

In the context of our genetic algorithm, we introduce some possibility of \textit{mutation of strategies}. With probability $\rho$, the child strategy mutates. That is, this element is added or subtracted with equal probability $\varepsilon$. The choice of the mutation probability ideally allows the model to explore sufficient diversity, while not generating excessive noise and falling to a quasi random search. Selecting $\rho = 0.1$ is one judicious way to ensure that in average, one child element of the $10$-dimensional vector is going to be mutated. Likewise, we desire mutations to have an impact on strategy performance, without generating too severe disturbances, a choice currently achieved by setting $\varepsilon = 0.1$. Section \ref{calibration} proceeds to the calibration of these parameters. \\

Let us highlight our introduction of \textit{elitism} in the genetic algorithm process. We indeed select the absolute best strategy obtained at any period $t$, and insert it in the next generation of strategies $\textbf{S}_{t+1}$ without any mutation. This specification ensures that we are not improperly discarding a possibly optimal strategy.

\subsection{Iterating to improve strategy pool quality}

Once $9$ child strategies, and the absolute best strategy have been added to the vector $\textbf{S}$, now of dimension $20$, we delete the last generation, likely performing worse since this new generation is only composed of -possibly mutated- recombinations of weakly better strategies. If the worst strategies are not disregarded, the size of the strategy sets increases by a factor of $2$ at each iteration, rendering the problem computationally extremely costly after only a few iterations. Keeping these strategies would also bias the identification of the best strategies: maintaining poor strategies improperly increases the payoff of sub-optimal strategies. By forgetting worst strategies, we keep the problem computationally efficient, provide our genetic algorithm with quality training set, and ensure a positive evolution of the strategy space towards optimal strategies. Some potentially good strategies -e.g. adequate in the long run, but too sophisticated for now- might indeed be lost in the process: the mechanism of mutation of strategies contributes to maintain such diversity.\\

The resulting vector $\textbf{S}_{t+1}$ at the beginning of the next period is likely to contain weakly better strategies than the previous ones at $t$, since it is composed of weakly dominating strategies, and some of their averages. We iterate this full procedure of evaluation of strategies, generation of new candidates through genetic crossover and mutation, and consider the resulting matrix of strategies. We desire to achieve convergence to a Nash equilibrium of the $p$-guessing game with analogy partitions and unstable environments.

\section{Algorithm calibration}
\label{calibration}

A required task for the calibration of our genetic algorithm is the magnitude of mutations, i.e. local changes to the strategies aiming at exploring the sample space to improve the efficiency of the solutions identified, while not disturbing the learning process with an excessive amount of noise. 

\subsection{Mutation magnitude $\varepsilon$}

\begin{figure}[H]
  \begin{minipage}[b]{0.5\textwidth}
    \includegraphics[width=\textwidth]{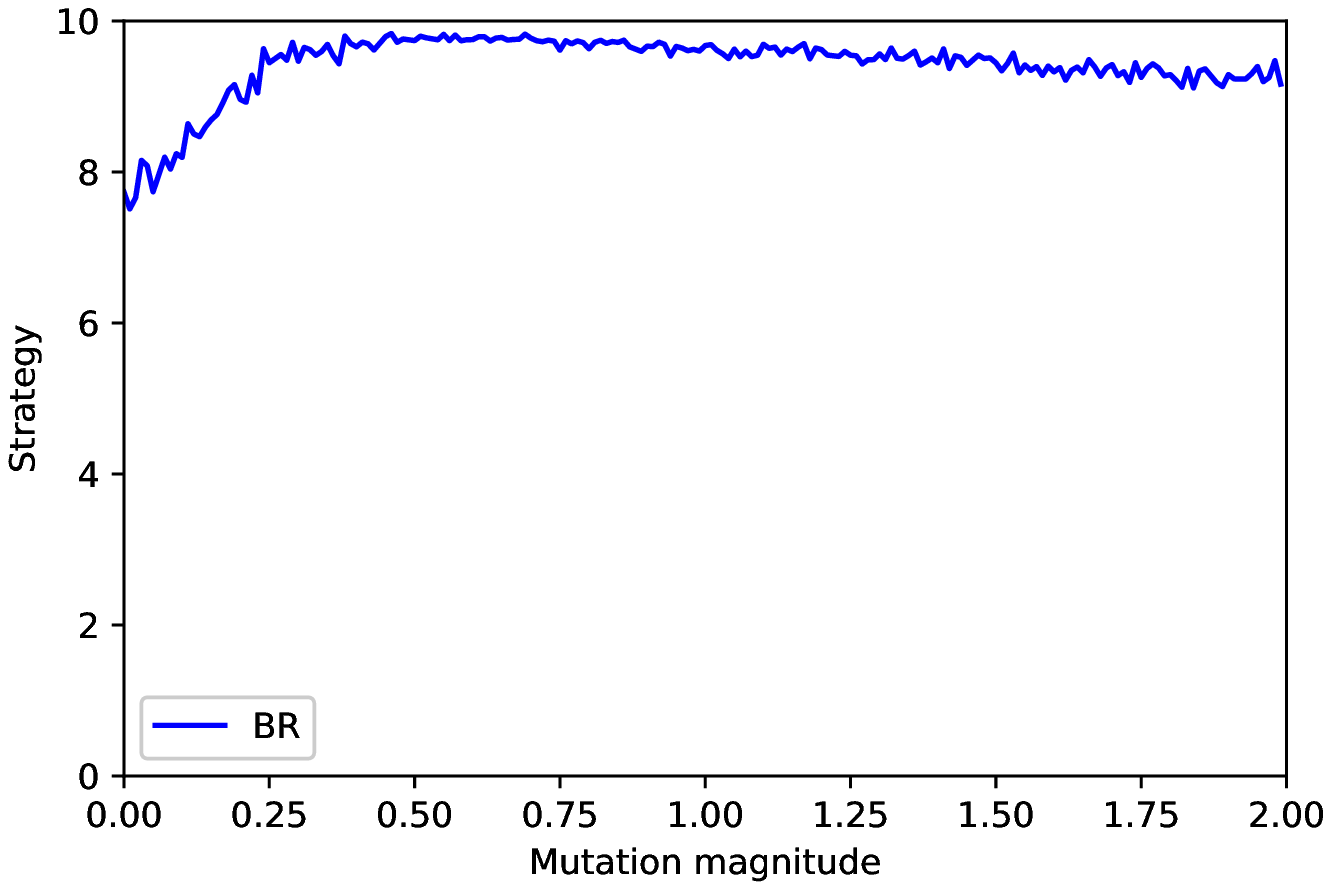}
    \caption{$q = 0$}
    \label{mutation0}
  \end{minipage}
  \begin{minipage}[b]{0.5\textwidth}
    \includegraphics[width=\textwidth]{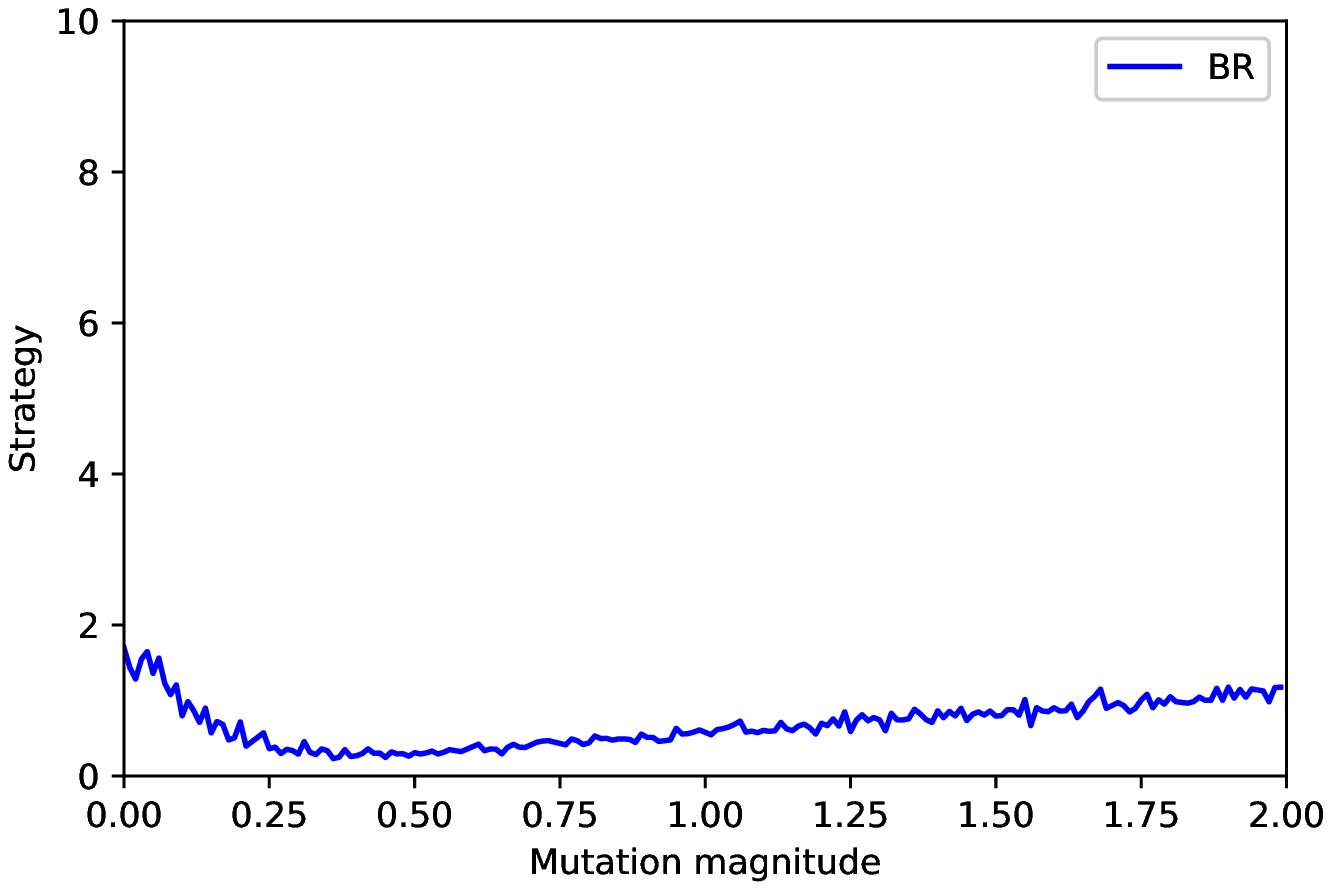}
    \caption{$q = 1$}
    \label{mutation1}
  \end{minipage}
\end{figure}

We test the model resulting strategy with various levels of mutations in well known stable environments, namely when $p > 1$ (Figure \ref{mutation0} where $q = 0$), and $p < 1$ (Figure \ref{mutation1} where $q = 1$), under a mutation probability of $\frac{1}{10}$ \textbf{ensuring that in average, one child strategy varies}. Simulations for each parameter combination are ran a hundred times to ensure robustness of the results. Evidently, when mutations are too low, in the range from 0 to around $0.4$, local perturbations of strategies are not sufficient to offer a better performance, and the algorithm converges to sub-optimal outcomes with respect to Nash equilibria. When mutations become too large, inefficiencies in outcomes are also observed. For the forthcoming results, we hence adopt a mutation amplitude that minimizes deviations from Nash equilibrium in these known cases, that is, $0.5$.

\subsection{Mutation probability $\rho$}

Likewise, setting the mutation magnitude to the identified level of $0.5$, we test various values for the mutation probabilities in the same well known cases to calibrate our model parameters. Simulations for each parameter combination are ran a hundred times to ensure robustness of the results. Absence, and to a lower extent, excessive probability of mutation of strategies generates deviations from the Nash equilibrium in these cases. We adopt a mutation probability of $0.1$, that constitutes an adequate parameter to reduce such inefficiencies of learning. The choice of our couple of parameters with mutation magnitude $\varepsilon$ of approximately $0.5$ and probability $\rho$ of $0.1$ is robust to brute force optimization methods, testing each possible combination of parameters.

\begin{figure}[H]
  \begin{minipage}[b]{0.5\textwidth}
    \includegraphics[width=\textwidth]{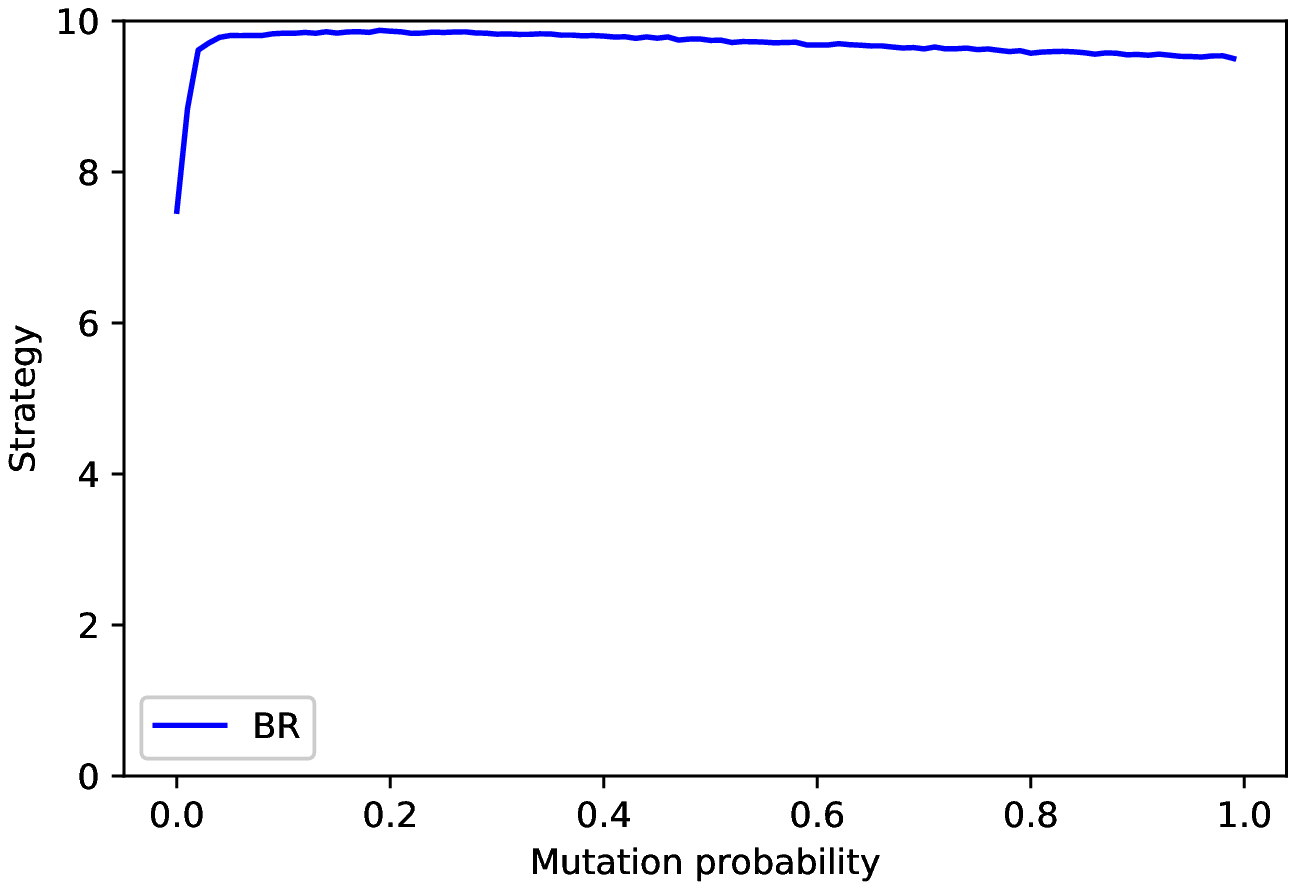}
    \caption{$q = 0$}
    \label{mutation_proba0}
  \end{minipage}
  \begin{minipage}[b]{0.5\textwidth}
    \includegraphics[width=\textwidth]{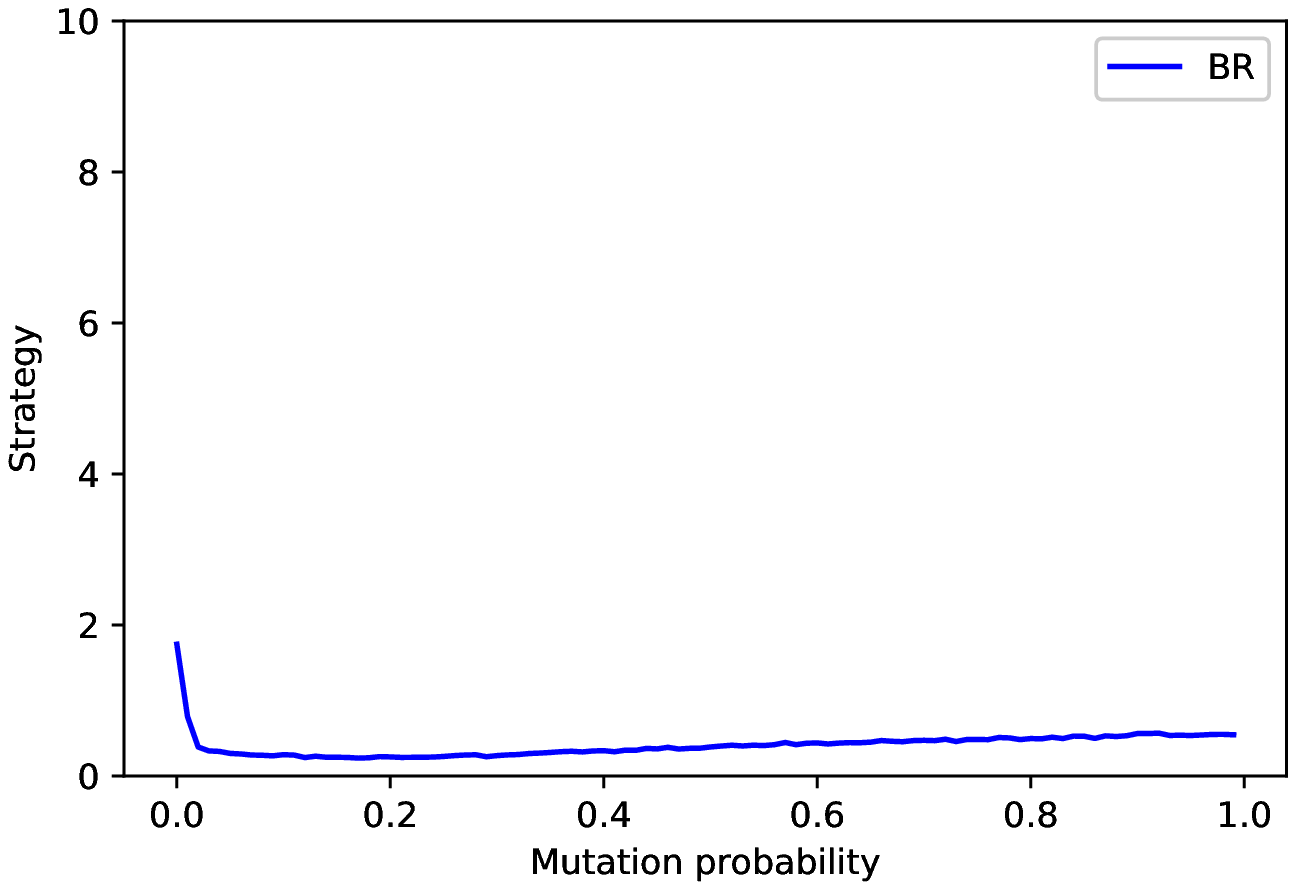}
    \caption{$q = 1$}
    \label{mutation_proba1}
  \end{minipage}
\end{figure}

Note that this calibration exercise is valid for the study of the range of actions $(0,10)$. For a different action range, say, higher ($[0,100]$), it is likely that either mutation probability and amplitude parameters, or both, should be increased to be adequate to the exploration of such a much larger strategy space. 

\section{Results}
\label{results}

\subsection{Convergence of learning with analogy-partitions in persistent environments}

\begin{figure}[H]
    \centering
    \includegraphics[scale = 0.65]{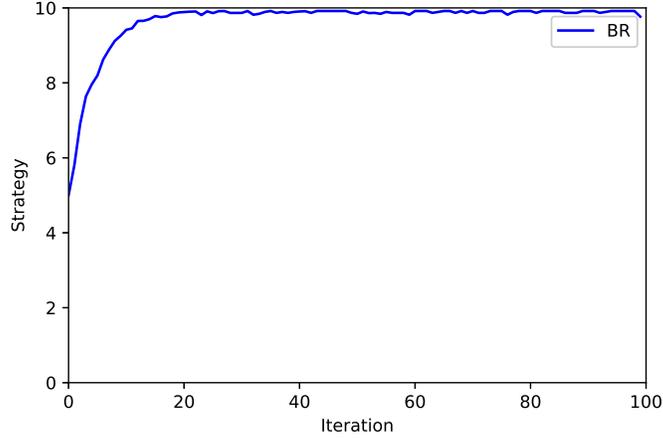}
    \caption{Average strategy across 100 model runs, $q = 0$}
    \label{convergence0}
\end{figure}

\begin{figure}[H]
    \centering
    \includegraphics[scale = 0.65]{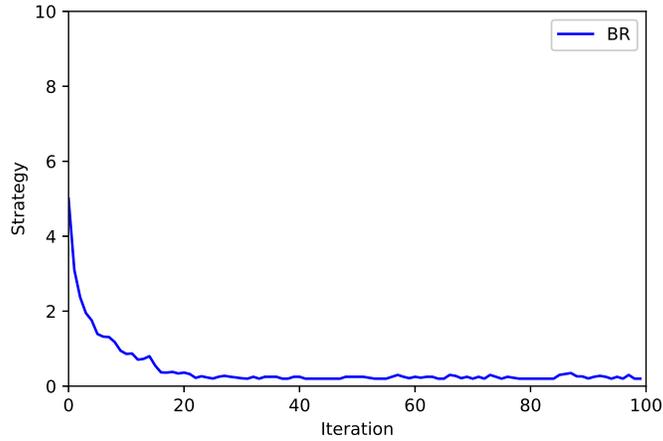}
    \caption{Average strategy across 100 model runs, $q = 1$}
    \label{convergence1}
\end{figure}

As a validation test of our calibrated learning model, we study the behavior of the model over the number of iterations of the genetic algorithm. We test two well known situations for which the Nash equilibrium is known to ensure our model displays adequate convergence. Simulations for each parameter combination are ran a hundred times to ensure robustness of the results. Figure \ref{convergence0} studies the case in which $q = 0$, namely all trials faced by the agents are such that $p > 1$. Conversely, Figure \ref{convergence1} focuses on trials with $p < 1$. Strategies identified by the evolutionary algorithm convergence approximately to the Nash equilibrium after 20 iterations in average, and displays stability once these outcomes have been obtained. The resulting convergence curves of Figures \ref{convergence0} and \ref{convergence1} are quite similar to the experimental results on evolution of strategies of \textcolor{blue}{Nagel (1995)} and \textcolor{blue}{Duffy and Nagel (1997)}.

\subsection{Learning with analogy partitions in inconstant $p$-guessing games}

When we depart from well-known cases to allow the value of $q$ to be in the interior interval $]0,1[$, we consider a situation in which an economic agent with learning abilities in the context of our evolutionary process, faces successive $p$-guessing games, that are bundled together in an analogy partition. As outlined above, the parameter $q$, probability than a game faced by the agent will have a parameter $p$ inferior to $1$, will impact the strategies that are inferred to be successful. When $q = 0.1$ for example, most games will have a $p$ superior to $1$, generating convergence towards the highest announcement in the action range. But occurrence of sporadic games in which $p < 1$ will impact this learning process. 

\begin{figure}[H]
    \centering
    \includegraphics[scale = 0.65]{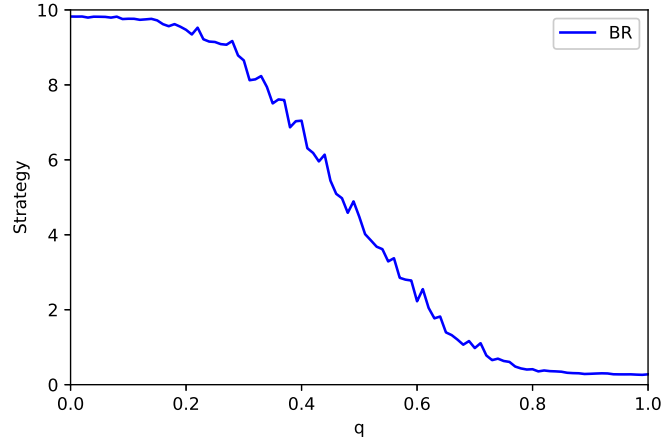}
    \caption{Average strategies with $q$ after 100 iterations across 100 trials}
    \label{analogyq}
\end{figure}

\begin{figure}[H]
    \centering
    \includegraphics[scale = 0.65]{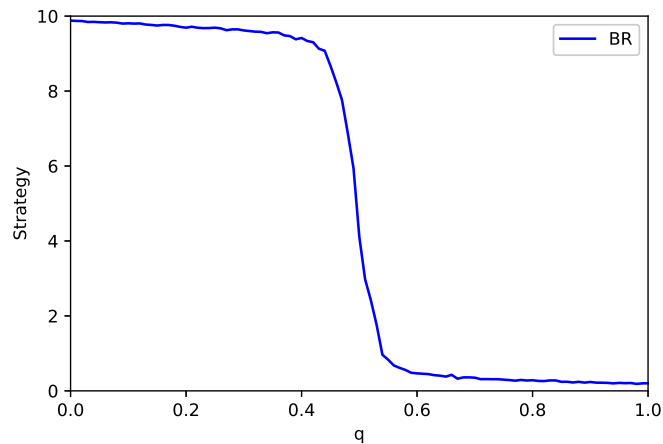}
    \caption{Average strategy after 1000 iterations across 100 trials}
    \label{analogyq_1000}
\end{figure}

Figure \ref{analogyq} show our results of the evolutionary process of learning for these interior values of $q$, when we give the program $100$ iterations. In line with Figures \ref{convergence0} and \ref{convergence1}, as well as theoretical and experimental results (\textcolor{blue}{Nagel (1995), Duffy and Nagel, 1997}), Nash equilibria are identified by the algorithm for the boundaries of the interval in which $q = \{0,1\}$. In between, convergent strategies take the shape of an inverted S-logistic curve, showing non-linearity in the way the probability distribution in the state of nature ($q$) impacts the learning process. How non-linear this process is, depends on the number of iterations we endow our genetic algorithm with. With $1000$ iterations, the program gives rises to strategies that are strongly non-linear with respect to $q$ represented in Figure \ref{analogyq_1000}, and correspond to playing the Nash equilibrium in almost all situations, regardless of the bundling of experiences.\\

\begin{figure}[H]
    \centering
    \includegraphics[scale = 0.65]{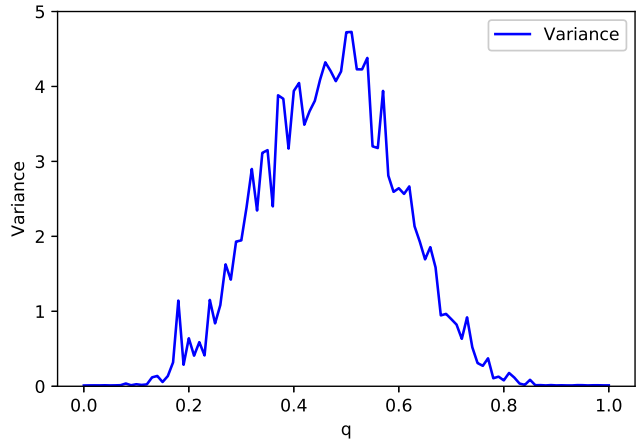}
    \caption{Variance of the average strategy after 100 iterations across 100 trials}
    \label{varianceq}
\end{figure}

When $q$ is close to $0$ (or $1$), the learning process is only marginally affected by occurrences of unexpected states of the world. Bundling previous experiences into a single analogy partition, and best responding to the expectations formed by experiences, is not perfectly linear with respect to probability distributions of the state of the world. Although for the perfectly mixed case ($q = \frac{1}{2}$), aggregating the strategies obtained across a hundred runs of the model yields a middle solution in the action range, this aggregation covers high volatility in such case, as shown in figure \ref{varianceq}. While interval boundaries, i.e. $q \in [0,0.2]$ and $q \in [0.8,1.0]$ show quasi null variance of strategies across model runs at $100$ iterations, interior cases give rise to a much richer set of strategies, and allow various trajectories. This diversity is likely to emerge from possibly different successive realizations of the draws of the values of $p$. In this stochastic process, drawing consecutive values of $p$ say inferior to $1$ is likely to shape the learning outcome. 

\begin{figure}[H]
    \centering
    \includegraphics[scale = 0.65]{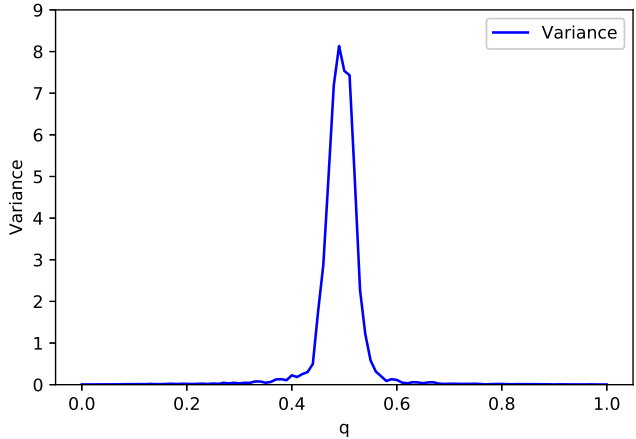}
    \caption{Variance of the average strategy after 1000 iterations across 100 trials}
    \label{varianceq1000}
\end{figure}

Increasing the number of iterations from 100 to 1000 as shown in figure \ref{varianceq1000} does not allow to completely erase this volatility of strategies for (near) mixed cases. However, the noise on strategies can be considerably reduced for larger intervals in the values of $q$, namely $q \in [0,0.4]$ and $q \in [0.6,1.0]$. It is likely that as the number of iterations grows to infinity, convergence may be obtained for any value of $q$ but $\frac{1}{2}$ that may constitute a perfectly random case, and only appear to correspond to playing the median of the action range by aggregating over repeated runs of the model.

\section{Endogenizing level-$k$ players in $p$-guessing games}
\label{discussion}

In the eyes of an external agent witnessing the strategies played by our evolutionary agent learning with analogy partitioning across inconstant situations with only $100$ (Figure \ref{analogyq}) attempts may seem random, or irrational. After learning about the game in a changing environment, e.g. with $q = \frac{1}{2}$, the agent learned to best respond when the $p$-guessing game is as likely to have $p < 1$ than the converse. If this agent was to play the evolutionary strategy he learned to be best across this training period in a game in which $p$ is, say, inferior to $1$, this agent would play randomly, or uniformly only in expectation, in the eyes of an external observer. Indeed, while averaging over model runs, the strategy played after learning in a $q = \frac{1}{2}$ environment is to play the middle of the action range (Figure \ref{analogyq}), there was high variance of the strategies played in this configuration. Even though this variance is rationalizable taking bounded rationality and the instability of the learning environment, it naturally appears random to the eyes of a more sophisticated another agent. \\

While the interplay of depth of reasoning types in $p$-guessing games is well understood in boundaries of the action range, little has been studied about the origin of these types in the first place, and how they may fit into a learning interpretation of the behavior of rational players. Considering low level players as bounded rational agents, that based their learning of the game and identification of optimal strategies on learning in a non-persistent environment under a relatively low ($100$) number of periods, allows to reconcile their apparent irrational behavior with the learning interpretation in the line of which the existence of such players is not straightforward.\\

\begin{figure}[H]
    \centering
    \includegraphics[scale = 0.65]{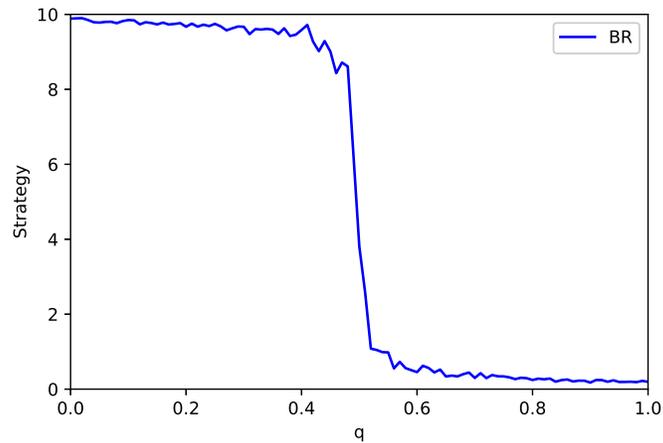}
    \caption{Average strategy after 10 iterations across 5000 trials}
    \label{analogyq_5000}
\end{figure}

In the continuity of this reasoning, plays with higher levels of depth of reasoning remain based on an expectation about the average behavior of other players. Either assuming the average plays according to the expectation of an uniform distribution, or according to a first-order best response to this anticipation, these beliefs about the average behavior of other players fit in this learning view. While Figure \ref{analogyq} showed the result of the learning process after $100$ iterations, pursuing a higher number of iterations, i.e. giving our agent more time and more experience to figure out how to play the game optimally, Figure \ref{analogyq_1000} replicates the behavior of a higher-order players in terms of reasoning. Continuing the process from the strategy shown in Figure \ref{analogyq} is indeed analog to having an agent best responding to such strategy. As an illustration, Figure \ref{analogyq_1000} below corresponds to the behavior of an agent with a much higher level of reasoning, with strong non-linearity in the response function. Conversely, Figure \ref{analogyq_10} of the supplementary material section shows the behavior of the agent after only $10$ iterations, much more linear and closer to the pure random play achieved with a null number of iterations. The agent after $1000$ iterations of our evolutionary process with analogy partitions plays the Nash equilibrium for almost all values of $q$. The shape of best responses varies little for more than 1000 iterations, as outlines Figure \ref{analogyq_5000}. The variance of strategies, i.e. convergence of the algorithm for interior (but not perfectly around $\frac{1}{2}$) values of $q$ illustrated in Figure \ref{varianceq1000} shows that this response displays stability. 

\section{Conclusion}
\label{conclusion}

We introduce an evolutionary process of learning to investigate the dynamics of learning and the resulting optimal strategies in $p$-guessing games environments with analogy partitions to understand how analogy reasoning impacts strategic behavior notably in unstable environments where the value of $p$ is not persistent through time, and depends on the realization of a state of nature. Our genetic algorithm behaves consistently with theoretical and experimental results in persistent environments with fixed $p$, converging to the Nash equilibrium after a small number of iterations. We explore strategic behavior learned in $p$-guessing games with varying values of $p$. Varying the number of iterations given to the genetic algorithm to learn about the game replicates the behavior of agents with different levels of depth of reasoning in the level $k$ approach. This evolutionary approach hence proposes a learning foundation for endogenizing existence and transitions between levels of reasoning in cognitive hierarchy models, consistent with previous theoretical and experimental results.

\section*{Compliance with Ethical Standards}
\textbf{Funding}: the author has no funding to report.\\
\textbf{Conflict of Interest}: The author declare that he has no conflict of interest.

\section*{References}
\begin{enumerate}
    \item Arthur, W. B. (1994). Inductive reasoning and bounded rationality. The \textit{American Economic Review}, 84(2), 406-411.
    \item Duffy, J. (2006). Agent-based models and human subject experiments. Handbook of computational economics, 2, 949-1011.
    \item Duffy J. \& Nagel R. (1997). On the robustness of behaviour in Experimental  `Beauty Contests' Games. \textit{The Economic Journal}, 107 (November), 1684-1700
    \item Edmonds, B., \& Moss, S. (1997). Modelling bounded rationality using evolutionary techniques. In \textit{AISB International Workshop on Evolutionary Computing} (pp. 31-42). Springer, Berlin, Heidelberg.
    \item Jehiel, P. (2005). Analogy-based expectation equilibrium. \textit{Journal of Economic theory}, 123(2), 81-104.
    \item Keynes, J. M. (1936). \textit{The general theory of employment, interest, and money} (ed. 2018). Springer.
    \item Friedman, D. (1998). On economic applications of evolutionary game theory. \textit{Journal of evolutionary economics}, 8(1), 15-43.
    \item Marks, R. E. (2002). Playing games with genetic algorithms. In \textit{Evolutionary computation in economics and finance} (pp. 31-44). Physica, Heidelberg.
    \item Mitchell, M. (1998). \textit{An introduction to genetic algorithms}. MIT press.
    \item Moulin, Herv\'{e} (1986). \textit{Game Theory for the Social Sciences} (2nd ed.). New York: NYU Press
    \item Nagel R. (1995). Unraveling in Guessing Games: An Experimental Study. \textit{American Economic Review}
    \item Riechmann, T. (2001). Genetic algorithm learning and evolutionary games. \textit{Journal of Economic Dynamics and Control}, 25(6-7), 1019-1037.
    \item Sbriglia, P. (2008). Revealing the depth of reasoning in p-beauty contest games. \textit{Experimental Economics}, 11(2), 107-121.
    \item Schelling, T. C. (1980). \textit{The strategy of conflict}. Harvard university press.
    \item Selten, R. and Stoecker, R. (1986. ‘End behaviour in sequences of finite prisoner’s dilemma supergames: a learning theory approach.’ \textit{Journal of Economic Behaviour and Organization}, vol. 7, no. 1, pp. 47-70
    \item Silver, D., Huang, A., Maddison, C. J., Guez, A., Sifre, L., Van Den Driessche, G., ... \& Dieleman, S. (2016). Mastering the game of Go with deep neural networks and tree search. \textit{Nature}, 529(7587), 484.
    \item Silver, D., Schrittwieser, J., Simonyan, K., Antonoglou, I., Huang, A., Guez, A., ... \& Chen, Y. (2017). Mastering the game of go without human knowledge. \textit{Nature}, 550(7676), 354.
    \item Silver, D., Hubert, T., Schrittwieser, J., Antonoglou, I., Lai, M., Guez, A., ... \& Lillicrap, T. (2018). A general reinforcement learning algorithm that masters chess, shogi, and Go through self-play. \textit{Science}, 362(6419), 1140-1144.
    \item Stahl, D. O. (1996). Boundedly rational rule learning in a guessing game. \textit{Games and Economic Behavior}, 16(2), 303-330.
    \item Vinyals, O., Babuschkin, I., Czarnecki, W. M., Mathieu, M., Dudzik, A., Chung, J., ... \& Oh, J. (2019). Grandmaster level in StarCraft II using multi-agent reinforcement learning. \textit{Nature}, 575(7782), 350-354.
    \item Weber, R. A. (2003). ‘Learning’with no feedback in a competitive guessing game. \textit{Games and Economic Behavior}, 44(1), 134-144.
\end{enumerate}

\section*{Supplementary Material}
\label{suppmat}

\subsection*{Additional iterations outcomes}

\begin{figure}[H]
    \centering
    \includegraphics[scale = 0.65]{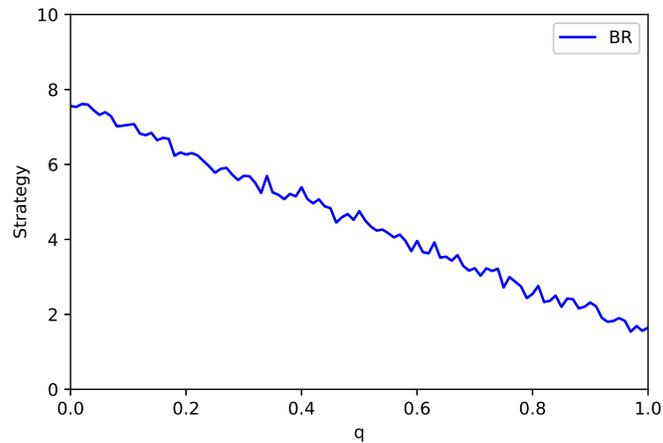}
    \caption{Average strategy after 10 iterations across 100 trials}
    \label{analogyq_10}
\end{figure}

\subsection*{Alternative payoff allocation mechanism}

\begin{figure}[H]
    \centering
    \includegraphics[scale = 0.65]{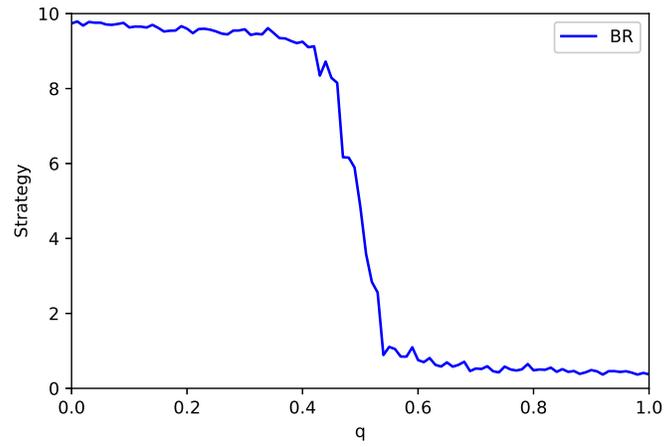}
    \caption{Average strategy after 10 iterations across 1000 trials with alternative payoff mechanism}
    \label{alt.1000}
\end{figure}

In order to ensure that the results above can to some extent be generalised to various settings of $p$-guessing games, we proceed to the same learning process with alternative specifications. We replicate below the setting chosen in most experiments. Payoffs (utilities) are not continuous: only the player with the best guess wins a prize. Other players do not win anything. As figure \ref{alt.1000} demonstrates, this setting still displays our results and the S-shaped best response function.

\end{document}